\documentclass[%
 reprint,
 superscriptaddress,
 amsmath,amssymb,
 aps,
floatfix,
]{revtex4-2}

\usepackage{graphicx}
\usepackage{dcolumn}
\usepackage{bm}

\usepackage[mathlines]{lineno}

\usepackage{color}
\usepackage[usenames,dvipsnames]{xcolor}

\usepackage{float}
\usepackage[colorlinks=true,
linkcolor=blue,citecolor=blue,
pdfauthor={ },
pdftitle={ },
pdfsubject={ },
pdfkeywords={ }]{hyperref}

\begin{document}


\title{Observation of Floquet topological phases with large Chern numbers}

\author{Kai Yang}
\thanks{These authors contributed equally to this work}
\affiliation{CAS Key Laboratory of Microscale Magnetic Resonance and School of Physical Sciences, University of Science and Technology of China, Hefei 230026, China}
\affiliation{CAS Center for Excellence in Quantum Information and Quantum Physics, University of Science and Technology of China, Hefei 230026, China}
\affiliation{Hefei National Laboratory, University of Science and Technology of China, Hefei 230088, China}

\author{Shaoyi Xu}
\thanks{These authors contributed equally to this work}
\affiliation{CAS Key Laboratory of Microscale Magnetic Resonance and School of Physical Sciences, University of Science and Technology of China, Hefei 230026, China}
\affiliation{CAS Center for Excellence in Quantum Information and Quantum Physics, University of Science and Technology of China, Hefei 230026, China}
\affiliation{Hefei National Laboratory, University of Science and Technology of China, Hefei 230088, China}

\author{Longwen Zhou}
\email{zhoulw13@u.nus.edu}
\thanks{These authors contributed equally to this work}
\affiliation{College of Physics and Optoelectronic Engineering, Ocean University of China, Qingdao 266100, China}

\author{Zhiyuan Zhao}
\affiliation{CAS Key Laboratory of Microscale Magnetic Resonance and School of Physical Sciences, University of Science and Technology of China, Hefei 230026, China}
\affiliation{CAS Center for Excellence in Quantum Information and Quantum Physics, University of Science and Technology of China, Hefei 230026, China}
\affiliation{Hefei National Laboratory, University of Science and Technology of China, Hefei 230088, China}

\author{Tianyu Xie}
\affiliation{CAS Key Laboratory of Microscale Magnetic Resonance and School of Physical Sciences, University of Science and Technology of China, Hefei 230026, China}
\affiliation{CAS Center for Excellence in Quantum Information and Quantum Physics, University of Science and Technology of China, Hefei 230026, China}
\affiliation{Hefei National Laboratory, University of Science and Technology of China, Hefei 230088, China}

\author{Zhe Ding}
\affiliation{CAS Key Laboratory of Microscale Magnetic Resonance and School of Physical Sciences, University of Science and Technology of China, Hefei 230026, China}
\affiliation{CAS Center for Excellence in Quantum Information and Quantum Physics, University of Science and Technology of China, Hefei 230026, China}
\affiliation{Hefei National Laboratory, University of Science and Technology of China, Hefei 230088, China}

\author{Wenchao Ma}
\affiliation{Department of Chemistry, Massachusetts Institute of Technology, Cambridge, Massachusetts 02139, USA}

\author{Jiangbin Gong}
\affiliation{Department of Physics, National University of Singapore, Singapore 117543}

\author{Fazhan Shi}
\affiliation{CAS Key Laboratory of Microscale Magnetic Resonance and School of Physical Sciences, University of Science and Technology of China, Hefei 230026, China}
\affiliation{CAS Center for Excellence in Quantum Information and Quantum Physics, University of Science and Technology of China, Hefei 230026, China}
\affiliation{Hefei National Laboratory, University of Science and Technology of China, Hefei 230088, China}

\author{Jiangfeng Du}
\email{djf@ustc.edu.cn}
\affiliation{CAS Key Laboratory of Microscale Magnetic Resonance and School of Physical Sciences, University of Science and Technology of China, Hefei 230026, China}
\affiliation{CAS Center for Excellence in Quantum Information and Quantum Physics, University of Science and Technology of China, Hefei 230026, China}
\affiliation{Hefei National Laboratory, University of Science and Technology of China, Hefei 230088, China}


\begin{abstract}
One of the most intriguing advantage of Floquet engineering is to generate new phases with large topological invariants. In this work, we experimentally simulate a periodically quenched generalized Haldane model on an NV center in diamond, and observe its Floquet Chern insulator phases with Chern numbers $C=1,2,4$ by imaging the static and dynamic spin textures in momentum space. Our work reveals the power of Floquet driving in transforming system's topology and generating large Chern number phases. It further establishes a unique experimental scheme to detect Floquet topological phases in two and higher spatial dimensions.
%
\end{abstract}

\maketitle

{\it Introduction}.--Floquet systems possess rich nonequilibrium topological phases triggered by time-periodic driving fields \cite{FTPRev1,FTPRev2,FTPRev3,FTPRev4}. They are featured with large topological numbers \cite{DerekPRL2012,TongPRB2013,ZhouPRA2018}, unique symmetry classifications \cite{Nathan2015,Potter2016,RoyPRB2017} and anomalous edge states with no static analog \cite{RudnerPRX2013,Titum2016,ZhouPRB2018}. The experimental observation of Floquet topological matter in cold atoms \cite{JotzuNat2014,Asteria2019,Wintersperger2020}, photonic setups \cite{KitagawaNat2012,RechtsmanNat2013,HuPRX2015}, solid state systems \cite{WangSci2013,McIver2020,Chen2021} etc. further enriches the means to realize, engineer, control and detect quantum materials from a dynamical perspective \cite{FTPRev5}.

The Haldane Chern insulator model describes noninteracting particles hopping in a honeycomb lattice and subject to staggered magnetic flux that breaks the time-reversal symmetry without introducing net magnetic fields \cite{Haldane1988}. It serves as a cornerstone in the conceptualization of topological insulators. Floquet driving fields further play a key role in the first realization of Haldane model by ultracold atoms \cite{JotzuNat2014}. However, assisted by high-frequency drivings \cite{OkaPRB2009}, the experimentally established Haldane model only exhibits phases with Chern numbers $C=\pm1$, which could not fully reveal the power of Floquet approach in realizing new states with large topological invariants. Meanwhile, theoretical studies suggest that Floquet Chern insulators with large Chern numbers could emerge when strong and near-resonant drivings are applied, which reshape the structure of underlying static systems in a non-perturbative manner \cite{DerekPRL2012,DerekPRB2014,ZhouEPJB2014,ZhouPRB2014,YapPRB2017,XiongPRB2016}. In this work, we experimentally realize a periodically quenched generalized Haldane model (GHM) on a single spin quantum system in diamond, and observe its large Chern number Floquet phases by measuring the spin textures in momentum space.

\begin{figure*}[htbp]
	\includegraphics[width=2\columnwidth]{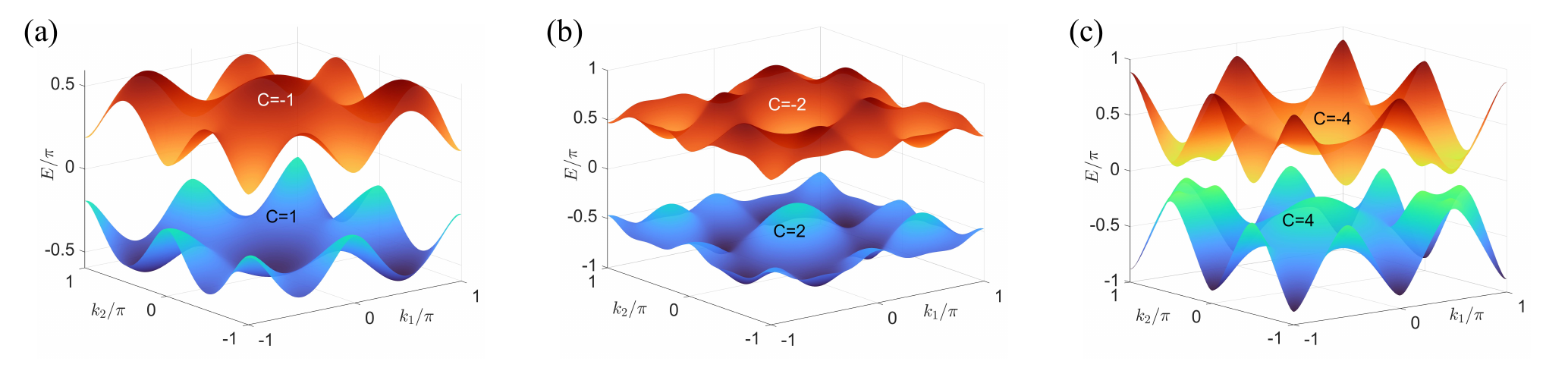}
	\caption{Quasienergy spectrum of the periodically quenched GHM under the PBC. $(T_1,T_2)$ are set as $(0.3,0.3)$, $(0.9,0.8)$ and $(0.9,1.2)$ in (a), (b), and (c), with Chern numbers marked on the two Floquet bands. Other parameters are $(t_1,t_2)=(1,0.8)$.}\label{theory}
\end{figure*}

\textit{Theory}.--We consider a generic two-dimensional two-band lattice model described by the Hamiltonian
$H=\sum_{{\bf k}\in{\rm BZ}}|{\bf k}\rangle H({\bf k})\langle{\bf k}|$ in momentum space, with $H({\bf k})={\bf h}({\bf k})\cdot\boldsymbol{\sigma}$.
Here ${\bf h}({\bf k})=(h_{x},h_{y},h_{z})$ is a three-component vector, 
$\boldsymbol{\sigma}=(\sigma_{x},\sigma_{y},\sigma_{z})$
are Pauli matrices, and ${\bf k}=(k_1,k_2)$ is the quasimomentum.
The Floquet operator of the system, which describes
its evolution over a driving period $T$ (e.g., from $t=0$ to $T$)
reads
\begin{equation}
U=\sum_{{\bf k}\in{\rm BZ}}|{\bf k}\rangle U({\bf k})\langle{\bf k}|,\,\,\, U({\bf k})=e^{-iT_{2}{\bf h}_{2}({\bf k})\cdot\boldsymbol{\sigma}}e^{-iT_{1}{\bf h}_{1}({\bf k})\cdot\boldsymbol{\sigma}}.\label{eq:U}
\end{equation}
Here we set the Planck constant $\hbar=1$ and utilize a piecewise quench protocol with the period $T=T_1+T_2$, in which ${\bf h}={\bf h}_1$ (${\bf h}={\bf h}_2$) for $t\in[0,T_1]$ ($t\in[T_1,T]$). The Floquet spectrum of the system is obtained by solving the eigenvalue equation $U({\bf k})|\varphi\rangle=e^{-iE({\bf k})}|\varphi\rangle$,
where the quasienergy $E({\bf k})\in[-\pi,\pi)$. In general, there are two quasienergy bands $E_{\pm}({\bf k})$ separated by gaps at $E=0,\pi$. Each gapped phase corresponds to a Floquet Chern insulator
(FCI) characterized by the Chern number
\begin{equation}
C=\int_{{\rm BZ}}\frac{d^{2}{\bf k}}{4\pi}\frac{{\bf d}({\bf k})\cdot\left[\partial_{k_{1}}{\bf d}({\bf k})\times\partial_{k_{2}}{\bf d}({\bf k})\right]}{\left|{\bf d}({\bf k})\right|^{3}}.\label{eq:CN}
\end{equation}
Here ${\bf d}({\bf k})=(d_{x},d_{y},d_{z})$ is a vector formed by the components
of the Floquet effective Hamiltonian ${\cal H}({\bf k})={\bf d}({\bf k})\cdot\boldsymbol{\sigma}$,
with ${\cal H}({\bf k})\equiv\frac{i}{T}\ln U({\bf k})$.
When the gap between $E_\pm({\bf k})$ closes/reopens at $E=0$ or $\pi$, the
system may undergo a topological phase transition followed by the
quantized jump of $C$ from one to another integer. 
Periodic driving fields could induce many such transitions and yield Floquet bands with large Chern numbers.

In experiments, we detect the Floquet Chern number of the system by imaging its spin texture in ${\bf k}$-space. Consider a general
normalized state $|\psi({\bf k})\rangle=\sum_{s=\pm}c_{s}({\bf k})|u_{s}({\bf k})\rangle$
with the quasimomentum ${\bf k}$, where $|u_{\pm}({\bf k})\rangle$
are Floquet eigenstates of ${\cal H}({\bf k})$ with quasienergies
$E_{\pm}({\bf k})$.
The expectation value of spin $\sigma_{j}$ over the state $|\psi({\bf k})\rangle$
is $\langle\sigma_{j}\rangle_{\bf k}=[|c_{+}({\bf k})|^{2}-|c_{-}({\bf k})|^{2}]d_{j}({\bf k})/E_{+}({\bf k})$.
The spin texture is then formed by the configuration of $(\langle\sigma_{i}\rangle_{\bf k},\langle\sigma_{j}\rangle_{\bf k},\langle\sigma_{l}\rangle_{\bf k})$
in ${\bf k}$-space for any $i,j,l=x,y,z$ with $i\neq j\neq l$.
Under the condition $|c_{+}({\bf k})|^{2}\neq|c_{-}({\bf k})|^{2}$,
the static winding angle (SWA) of the spin texture at each ${\bf k}$ is defined as
\begin{equation}
\theta_{jl}({\bf k})\equiv\arctan\left(\frac{\langle\sigma_{j}\rangle_{\bf k}}{\langle\sigma_{l}\rangle_{\bf k}}\right)=\arctan\left(\frac{d_{j}}{d_{l}}\right).\label{eq:SWA}
\end{equation}
Assuming $|c_{+}({\bf k})|^{2}>|c_{-}({\bf k})|^{2}$ without loss
of generality, we can extract the Chern number $C$ from the spin
texture $(\langle\sigma_{i}\rangle_{\bf k},\langle\sigma_{j}\rangle_{\bf k},\langle\sigma_{l}\rangle_{\bf k})$
through the relation \cite{ShenPRL2018,ZhuPRR2020},
\begin{equation}
C=\frac{1}{2}\sum_{{\bf k}_{0}\in{\rm SPs}}{\rm sgn}\left(\langle\sigma_{i}\rangle_{{\bf k}_{0}}\right)w({\bf k}_{0}).\label{eq:CN2}
\end{equation}
Each ${\bf k}_{0}$ resides at a phase singularity of $\theta_{jl}$
in ${\bf k}$-space, which appears under the condition $\langle\sigma_{j}\rangle=\langle\sigma_{l}\rangle=0$.
$w({\bf k}_{0})$ is an integer-quantized winding number defined along
an infinitesimal clockwise path $S_{{\bf k}_{0}}$ around ${\bf k}_{0}$, i.e.,
\begin{equation}
w({\bf k}_{0})=\oint_{S_{{\bf k}_{0}}}\frac{d{\bf k}}{2\pi}\partial_{{\bf k}}\theta_{jl}({\bf k}).\label{eq:WS}
\end{equation}
By preparing the system in the state $|\psi({\bf k})\rangle$
and measuring the spin texture $(\langle\sigma_{i}\rangle_{\bf k},\langle\sigma_{j}\rangle_{\bf k},\langle\sigma_{l}\rangle_{\bf k})$
in ${\bf k}$-space, the Chern numbers of different FCI phases in
the periodically quenched model can be obtained in experiments.

We can also extract the Chern number of FCIs from dynamic spin
textures \cite{ZhuPRR2020,ZhouPRB2019,ZhangPRL2020}. Let the system be prepared in the state $|\psi({\bf k})\rangle=\sum_{s=\pm}c_{s}({\bf k})|u_{s}({\bf k})\rangle$.
After the evolution over $\ell$ driving periods guided by $U({\bf k})$,
it reaches the state $|\psi({\bf k},\ell T)\rangle=\sum_{s=\pm}c_{s}({\bf k})e^{-i\ell TE_{s}({\bf k})}|u_{s}({\bf k})\rangle$,
yielding the spin expectation value $\overline{\sigma_{j}}({\bf k}, \ell T)=\langle\psi({\bf k},\ell T)|\sigma_{j}|\psi({\bf k},\ell T)\rangle$
for $j=x,y,z$. The long-time stroboscopic average of $\overline{\sigma_{j}}({\bf k},\ell T)$ is
\begin{equation}
\overline{\sigma_{j}}_{\bf k}=\frac{1}{NT}\sum_{\ell=1}^{N}\overline{\sigma_{j}}({\bf k},\ell T),\quad N\in\mathbb{Z},\quad N\gg1,\label{eq:SST}
\end{equation}
from which we obtain the dynamic winding angle 
\begin{equation}
\eta_{jl}({\bf k})=\arctan\left(\frac{\overline{\sigma_{j}}_{\bf k}}{\overline{\sigma_{l}}_{\bf k}}\right).\label{eq:DWA}
\end{equation}
In the limit $N\rightarrow\infty$ and under the condition $|c_{+}({\bf k})|^{2}>|c_{-}({\bf k})|^{2}$,
it can be shown that $\eta_{jl}({\bf k})=\theta_{jl}({\bf k})$ and
${\rm sgn}(\langle\sigma_{i}\rangle_{\bf k})={\rm sgn}(\overline{\sigma_i}_{\bf k}$),
with $i,j,l=x,y,z$ and $i\neq j\neq l$. This finally allows us to
obtain the Chern number $C$ in Eq.~(\ref{eq:CN2}) from the long-time
averaged dynamic spin texture $(\overline{\sigma_{i}}_{\bf k},\overline{\sigma_{j}}_{\bf k},\overline{\sigma_{l}}_{\bf k})$
in ${\bf k}$-space. In experiments, the dynamic approach allows one
to probe the nonequilibrium Floquet Chern topology without having
the complete knowledge about the Floquet Hamiltonian and the initial state.

\begin{figure}
	\includegraphics[width=1\columnwidth]{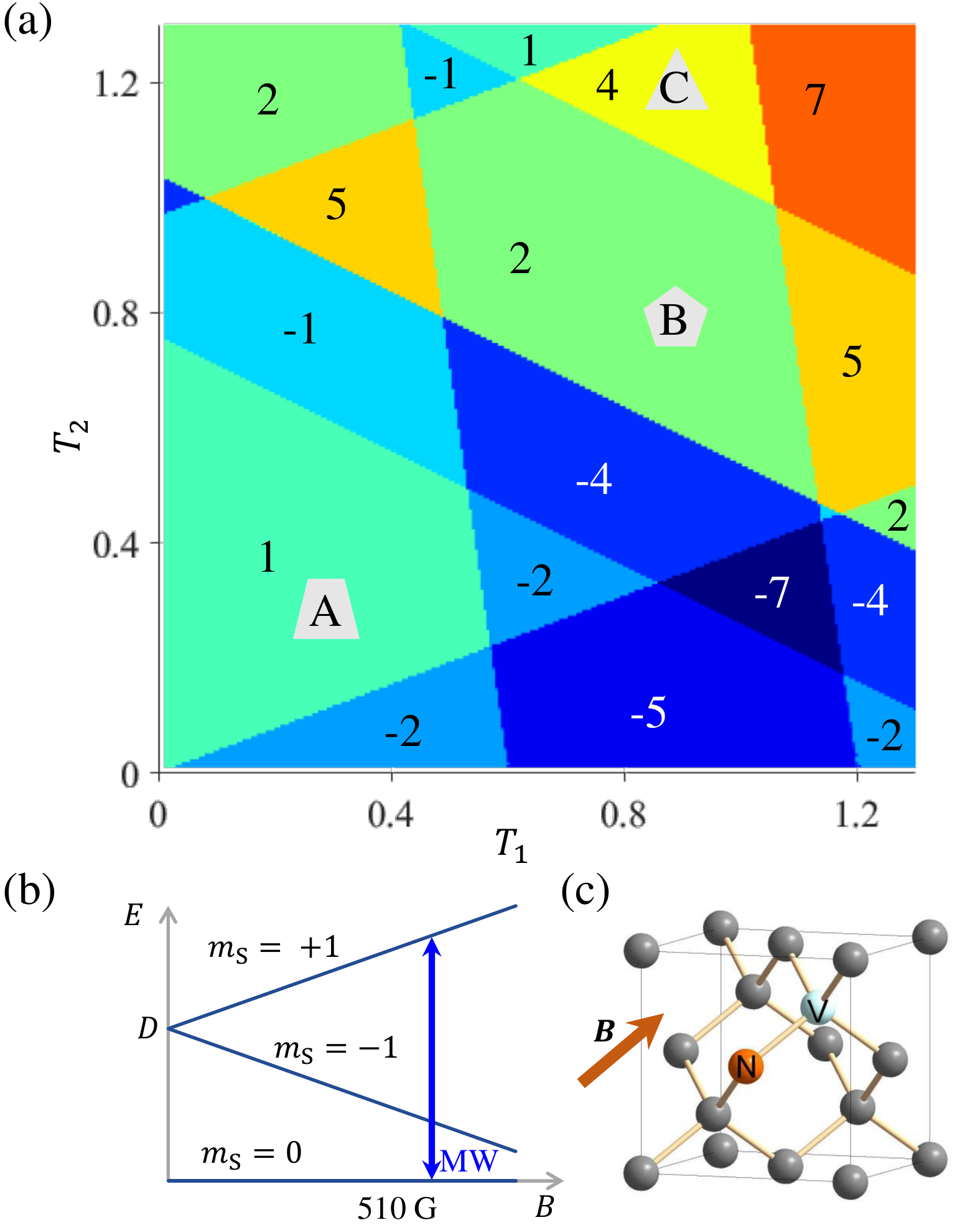}
	\caption{Topological phase diagram of the periodically quenched GHM and the experimental system.
(a) The topological phase diagram in $(T_1,T_2)$ space. Each area of the same color shares the same Chern number as marked therein. The three polygonal markers correspond to the three groups of parameters in our experiments. Other system parameters are $(t_1,t_2)=(1,0.8)$.
(b) Magnetic field dependence of the NV's energy levels in the ground state. A magnetic field of around 510 Gauss parallel to the NV axis is selected in the experiment.
(c) The structure of an NV center in the diamond lattice. The magnetic field is adjusted to be along the NV axis.
}\label{systerm}
\end{figure}

\textit{Model and Experiment}.--In the experiment, we realize the Bloch Hamiltonian of a GHM and apply time-periodic quenches to its hopping parameters. The components of GHM in ${\bf h}({\bf k})=(h_x,h_y,h_z)$ are explicitly given by
$h_{x}=t_{1}(1+\cos k_{1}+\cos k_{2})+t_{3}[2\cos(k_{1}-k_{2})+\cos(k_{1}+k_{2})]$,
$h_{y}=t_{1}(\sin k_{1}+\sin k_{2})+t_{3}\sin(k_{1}+k_{2})$, and
$h_{z}=2t_{2}\sin\phi[\sin k_{1}-\sin k_{2}-\sin(k_{1}-k_{2})]$.
$t_{1}$, $t_{2}$ and $t_{3}$ represent the first, second and third
neighbor hopping amplitudes. $\phi\in[-\pi,\pi)$ is a phase factor
accompanying the second neighbor hopping.
This GHM admits topological
phases with a maximal Chern number $C=2$ \cite{BenaPRB2011,SticletPRB2013}. Meanwhile, much richer
Chern topology could emerge via applying time-periodic quenches to the GHM
\cite{XiongPRB2016}. In this work, we realize piecewise periodic quenches
of the system parameters $(t_{3},\phi)$, so that $(t_{3},\phi)=(0.75,-\pi/6)$
for $t\in[nT,nT+T_{1})$ and $(t_{3},\phi)=(-0.75,-\pi/2)$ for $t\in[nT+T_{1},nT+T_{1}+T_{2})$.
Here $t$ denotes time, $n\in\mathbb{Z}$, and $T=T_{1}+T_{2}$ represents
the driving period. The system Hamiltonians in the time duration
$T_{1}$ and $T_{2}$ have the form of $H({\bf k})$ and differ only
in their parameters $(t_{3},\phi)$. Typical spectra
of the periodically quenched GHM are shown in Fig.~\ref{theory}, where we observe
two Floquet bands $E_{\pm}({\bf k})$ separated by quasienergy gaps at $E=0$
and $\pi$. The two bands could touch when \cite{XiongPRB2016}
\begin{alignat}{1}
& {\bf h}_{1}({\bf k})/|{\bf h}_{1}({\bf k})|= \pm{\bf h}_{2}({\bf k})/|{\bf h}_{2}({\bf k})|,\,\,\,{\rm and}\nonumber \\
& |{\bf h}_{1}({\bf k})T_{1}|\pm|{\bf h}_{2}({\bf k})T_{2}|= n\pi\,\,\,{\rm for}\,\,\,n\in\mathbb{Z}.\label{eq:PTC}
\end{alignat}
In Fig.~\ref{systerm}(a), we report the topological phase diagram of the periodically
quenched GHM for different quench durations $(T_{1},T_{2})$ \cite{XiongPRB2016}. Distinct
FCI phases carry different Chern numbers $C$ as defined in Eq.~(\ref{eq:CN}),
and phases with large $C$ ($=\pm4,\pm5,\pm7$) are observed, which go markedly
beyond the allowed Chern insulator phases in the non-driven Haldane model \cite{Haldane1988,BenaPRB2011,SticletPRB2013}.

We simulate the the periodically quenched GHM by a single spin system and measure its Floquet band Chern number in experiments based on a nitrogen-vacancy (NV) center in diamond \cite{SM}.
As shown in Fig.~\ref{systerm}{(c)}, the NV center \cite{doherty_NV_2013,schirhagl_NV_2014,wrachtrup_NV_2016} is a kind of impurity in the diamond crystal lattice.
It has emerged as one of the most promising systems for implementing quantum technologies due to its superior optical and spin properties \cite{MaPRL2018,YangPRB2019}. We use the spin state of NV to simulate the Floquet Hamiltonian of GHM in ${\bf k}$-space.
The NV center is addressed by a home-built confocal microscope. The $532$ nm laser which can be switched by an acousto-optic modulator (AOM) is focused into the diamond by an oil objective, and the excited fluorescence can be collected by the same oil objective and finally detected by an avalanche photodiode with a counter card. In our experiment, an external static magnetic field around $510$ G parallel to the NV symmetry axis [see Fig.~\ref{systerm}(b)] is used for polarization of the host $^{14}$N nuclear spin, owing to resonant polarization exchanges with the electronic spin in the excited state \cite{jacques_dynamic_2009}. The magnetic field is adjusted by a permanent magnet mounted on a three-axis translation stage.

The Hamiltonian of the NV center's electronic ground state with a magnetic field $B$ applied along the NV axis is $H_{\rm{NV}} = D S_z^2+\gamma B S_z$, where $S_z$ is the angular momentum operator for spin-$1$, $D=2\pi\times2870$~MHz is the zero-field splitting, and $\gamma=2\pi\times2.8$~MHz/G is the electron's spin gyromagnetic ratio. Here the NV axis is chosen to be the $z$-axis. We can choose an eigenstate $\left| {{m_s} =  +1} \right\rangle$ or $ \left| {{m_s} =  -1}\right\rangle$ to form a  well-defined qubit with $ | {{m_s} =  0}\rangle$. Driven by a designed microwave pulse sequence, the Hamiltonian of this subspace, which was employed for the simulation, can then be built using spin-$1/2$ operators. Here the microwaves driving the evolution of NV spin are generated from an arbitrary waveform generator (AWG), then enhanced by the power amplifier, and finally radiated to the NV center from a coplanar waveguide. All these microwave manipulations and qubit states are described in the rotating frame.

\begin{figure}
	\includegraphics[width=1\columnwidth]{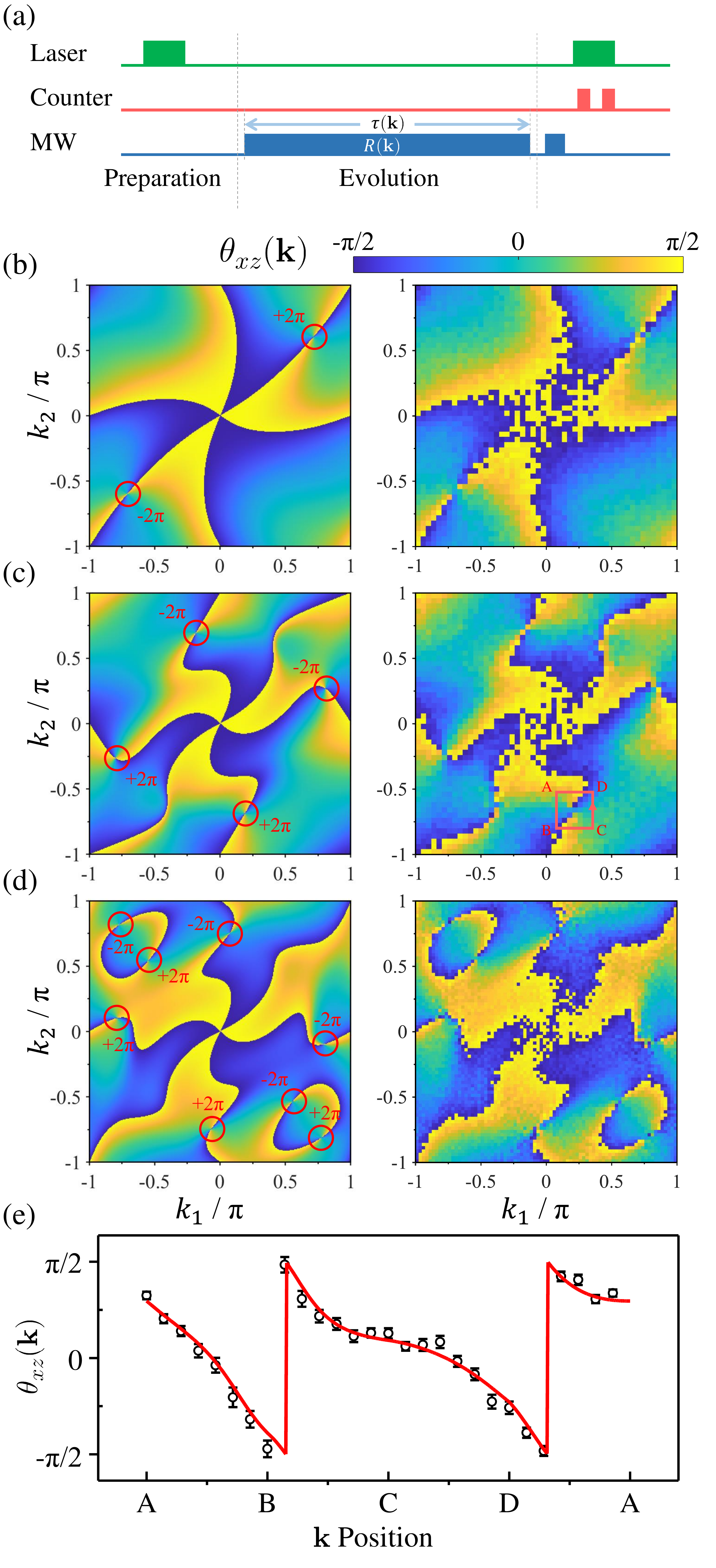}
	\caption{
Static winding angle (SWA). 
(a) Pulse sequence used for qubit control and measurement in SWA experiments.
(b)--(d) show the distributions of SWAs in ${\bf k}$-space, with the left and right panels obtained from theory and experiment, respectively.
Each singularity point is circled in red and marked with the corresponding winding angle. The parameters in (b), (c) and (d) are respectively set as $(T_1,T_2) = (0.3,0.3)$, $(0.9,0.8)$ and $(0.9,1.2)$. Other parameters are the same as those in Fig.~\ref{systerm}(a).
(e) The SWA distribution on a closed square shown in the right panel in (c). Black circles and the red curve represent experimental and theoretical results, respectively.}
	\label{SWA}
\end{figure}

In the SWA experiments shown in Fig.~\ref{SWA}, we run the pulse sequence at each pixel in ${\bf k}$-space separately.
Fig.~\ref{SWA}(a) shows the pulse sequence used in these experiments. This sequence is composed of three sections: preparation, evolution, and measurement. After going through the preparation section, the qubit state is initialized to $|0\rangle$ by a laser pulse. Then a resonant microwave pulse is applied to drive the state's evolution until the final state becomes an eigenstate of ${\cal H}(${\bf k}$)$. To derive the required driving pulse, we take every ${\bf k}$ as a control parameter to obtain the Hamiltonian ${\cal H}(${\bf k}$)$ and its eigenstates. The shape and duration of the pulse are denoted by $R({\bf k})$ and $\tau({\bf k})$ in Fig.~\ref{SWA}(a). The final section is used to measure $\sigma_x$, $\sigma_y$ or $\sigma_z$. For the measurement of $\sigma_x$ and $\sigma_y$, a $\pi/2$ rotation around the direction of $-y$ and $x$ respectively is implemented before the optical readout of NV spin state. This rotation pulse is not needed for $\sigma_z$. The final readout of the qubit is obtained by calculating the contrast of the photon count obtained by two counting windows, which correspond to the signal and reference detections. After repeating the sequence two hundred thousand times and taking the average, we get the value of $\left \langle \sigma _i \right \rangle_{\bf k}$ ($i =x,y,z$) at each ${\bf k}$.
Then we figure out the value of SWA following the Eq.~(\ref{eq:SWA}).

\begin{figure}
	\includegraphics[width=1\columnwidth]{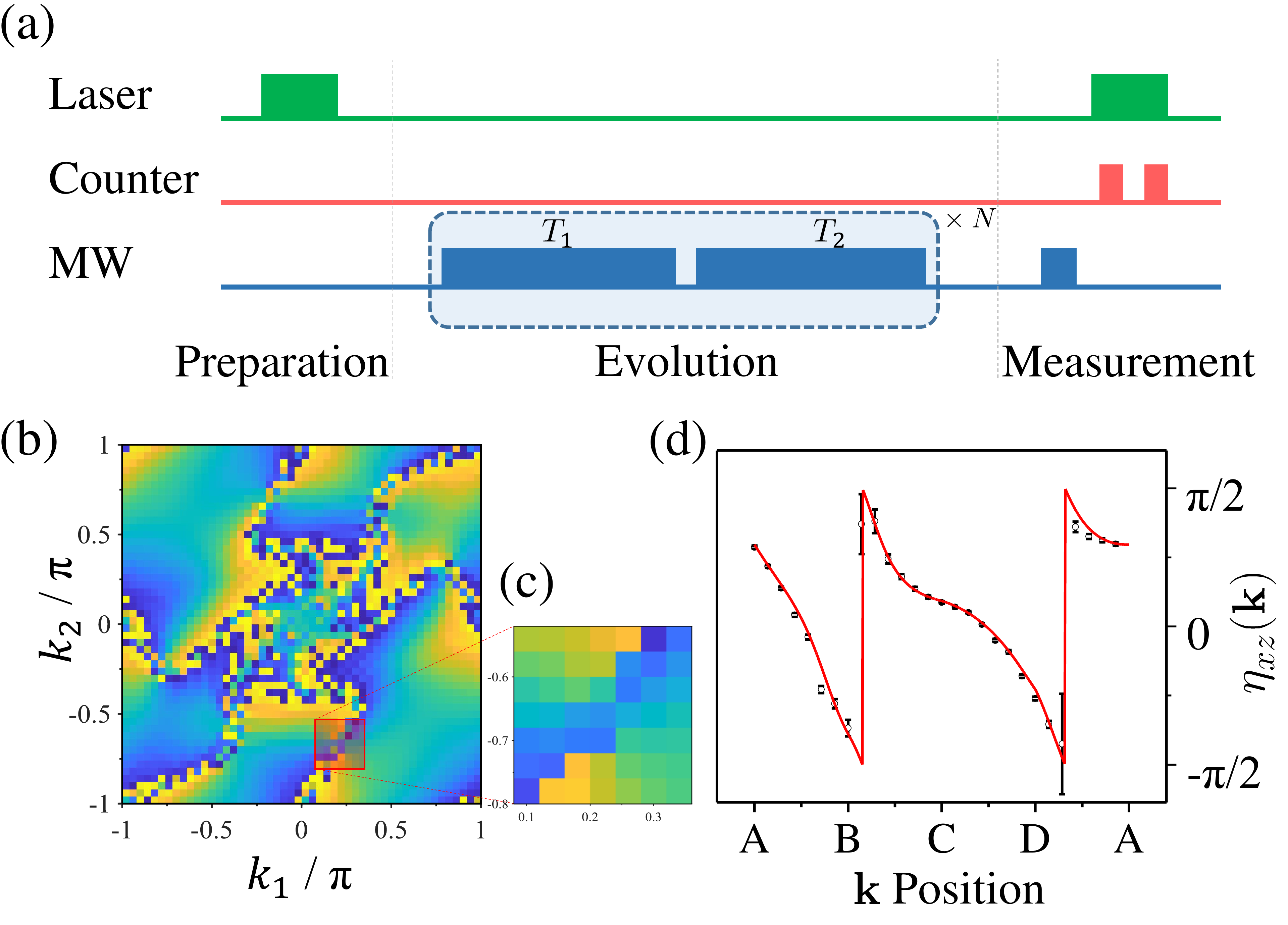}
	\caption{
Dynamical winding angle (DWA).
(a) Pulse sequence employed for qubit control and measurement in DWA experiments.
(b) The distribution of DWA in ${\bf k}$-space obtained from theoretical calculation. Here the system is evolved over $N=64$ driving periods under the time-periodic Hamiltonian whose parameters are chosen as $(T_1,T_2) = (0.9,0.8)$. The rest of the parameters are the same as those chosen for other experiments.
(c) Experimental demonstration of DWA. The value of ${\bf k}$ is selected from the marked boxed area in (b).
(d) DWA distribution on a closed loop depicted by the square boundary in (c). Black circles represent experimental values and the red curve shows the theoretical result from the SWA calculations, which are coincident with the DWA in the limit $N\rightarrow\infty$. The loop is the same as that used in Fig.~\ref{SWA}(c).
}
\label{DWA}
\end{figure}

In the experiments, we select three representative sets of parameters corresponding to different topological numbers in the phase diagram of the periodically quenched GHM, and detect the distributions of SWA in ${\bf k}$-space respectively. These selected parameters are marked in Fig.~\ref{systerm}(a) and correspond to the three spectra in Fig.~\ref{theory}. Figs.~\ref{SWA}(b)--(d) show the SWA distributions in ${\bf k}$-space obtained theoretically and experimentally.
Comparing the left and right panels in Figs.~\ref{SWA}(b)--(d), it is clear that our experimental results match the theoretical expectations. The detection of SWA near each singularity point is crucial to determine the Chern numbers of Floquet bands. In Fig.~\ref{SWA}(e), we show the SWA distribution along a closed square marked in Fig.~\ref{SWA}(c) to demonstrate the typical singularity with winding number $+1$. Following the Eqs.~(\ref{eq:CN2}) and (\ref{eq:WS}), one can calculate the Chern number using the winding numbers of all singularity points. The weight of each winding number is determined by the sign of $\left \langle \sigma _i \right \rangle_{\bf k}$ which is not used in the calculation of $\theta_{jl}({\bf k})$. The Chern numbers of ${\cal H}(${\bf k}$)$ are found to be $1$, $2$, and $4$ in the three groups of experiments.

Beyond the SWA, one can also obtain the Chern number of the system by probing the dynamic winding angle (DWA), which is robust to initial state preparations. We theoretically calculate the distribution of DWA and show it in Fig.~\ref{DWA}(b). The experimental results are shown in Fig.~\ref{DWA}(c), and the scanning region of the experiment is the area marked in Fig.~\ref{DWA}(b). We show the DWA distribution on the anticlockwise boundary of this area in Fig.~\ref{DWA}(d), in order to illustrate the singularity with winding number $+1$. 
The parameters adopted here are consistent with those in Fig.~\ref{SWA}(c) and the state is evolved over $N=64$ driving periods. The pulse sequence for measuring spin expectation values $\overline{\sigma_{j}}_{\bf k}$ ($j=x,y,z$) is shown in Fig.~\ref{DWA}(a).  
In the preparation, the spin can be initialized to an arbitrary state. For convenience, the initial state of each experiment is set to $|0\rangle$. During the evolution, the spin state is continually driven for $N=64$ periods, each of which includes two parts that corresponded to parameters $T_1=0.9$ and $T_2=0.8$, respectively. The measurement section of the experiment is the projection readout of the spin state on the three coordinate axes, which is the same as the last section in Fig.~\ref{SWA}(a).

{\it Summary}.--In this work, we experimentally realized a periodically quenched generalized Haldane model on an NV center in diamond, and observed FCI phases therein with large Chern numbers by imaging the static and dynamic spin textures in the momentum space. Our approach is generic, insensitive to the initial state preparation, and in principle extendable to the realization and detection of FCI phases following arbitrary driving schemes in two-band models. In future work, it would be interesting to consider generalizing our strategy to realize and detect Floquet topological phases in multi-band systems, higher spatial dimensions and other symmetry classes.

\textit{Acknowledgments}.--This work is supported by the National Key R\&D Program of China (Grant No.~2018YFA0306600), the National Natural Science Foundation of China (Grant No.~11905211, 81788101, T2125011), the CAS (Grant No.~XDC07000000, GJJSTD20200001, Y201984), Innovation Program for Quantum Science and Technology (Grant No.~2021ZD0302200, 2021ZD0303204), the Young Talents Project at Ocean University of China (Grant No.~861801013196), the Anhui Initiative in Quantum Information Technologies (Grant No.~AHY050000), the Applied Research Project of Postdoctoral Fellows in Qingdao (Grant No.~861905040009), Hefei Comprehensive National Science Center, and the Fundamental Research Funds for the Central Universities.

\nocite{*}
\bibliography{citing}

\begin{thebibliography}{10}
\bibitem{SM}
\bibinfo{title}{See {S}ec.~{I} in the {S}upplemental {M}aterial for details
  about the connection between our model and the manipulation of the {NV}
  qubit.}
\end{thebibliography}

\end{document}



\title{Supplemental Material for \protect\\Observation of Floquet topological phases with large Chern numbers}

\author{Kai Yang}
\thanks{These authors contributed equally to this work}
\affiliation{CAS Key Laboratory of Microscale Magnetic Resonance and School of Physical Sciences, University of Science and Technology of China, Hefei 230026, China}
\affiliation{CAS Center for Excellence in Quantum Information and Quantum Physics, University of Science and Technology of China, Hefei 230026, China}
\affiliation{Hefei National Laboratory, University of Science and Technology of China, Hefei 230088, China}

\author{Shaoyi Xu}
\thanks{These authors contributed equally to this work}
\affiliation{CAS Key Laboratory of Microscale Magnetic Resonance and School of Physical Sciences, University of Science and Technology of China, Hefei 230026, China}
\affiliation{CAS Center for Excellence in Quantum Information and Quantum Physics, University of Science and Technology of China, Hefei 230026, China}
\affiliation{Hefei National Laboratory, University of Science and Technology of China, Hefei 230088, China}

\author{Longwen Zhou}
\email{zhoulw13@u.nus.edu}
\thanks{These authors contributed equally to this work}
\affiliation{College of Physics and Optoelectronic Engineering, Ocean University of China, Qingdao 266100, China}

\author{Zhiyuan Zhao}
\affiliation{CAS Key Laboratory of Microscale Magnetic Resonance and School of Physical Sciences, University of Science and Technology of China, Hefei 230026, China}
\affiliation{CAS Center for Excellence in Quantum Information and Quantum Physics, University of Science and Technology of China, Hefei 230026, China}
\affiliation{Hefei National Laboratory, University of Science and Technology of China, Hefei 230088, China}

\author{Tianyu Xie}
\affiliation{CAS Key Laboratory of Microscale Magnetic Resonance and School of Physical Sciences, University of Science and Technology of China, Hefei 230026, China}
\affiliation{CAS Center for Excellence in Quantum Information and Quantum Physics, University of Science and Technology of China, Hefei 230026, China}
\affiliation{Hefei National Laboratory, University of Science and Technology of China, Hefei 230088, China}

\author{Zhe Ding}
\affiliation{CAS Key Laboratory of Microscale Magnetic Resonance and School of Physical Sciences, University of Science and Technology of China, Hefei 230026, China}
\affiliation{CAS Center for Excellence in Quantum Information and Quantum Physics, University of Science and Technology of China, Hefei 230026, China}
\affiliation{Hefei National Laboratory, University of Science and Technology of China, Hefei 230088, China}

\author{Wenchao Ma}
\affiliation{Department of Chemistry, Massachusetts Institute of Technology, Cambridge, Massachusetts 02139, USA}

\author{Jiangbin Gong}
\affiliation{Department of Physics, National University of Singapore, Singapore 117543}

\author{Fazhan Shi}
\affiliation{CAS Key Laboratory of Microscale Magnetic Resonance and School of Physical Sciences, University of Science and Technology of China, Hefei 230026, China}
\affiliation{CAS Center for Excellence in Quantum Information and Quantum Physics, University of Science and Technology of China, Hefei 230026, China}
\affiliation{Hefei National Laboratory, University of Science and Technology of China, Hefei 230088, China}

\author{Jiangfeng Du}
\email{djf@ustc.edu.cn}
\affiliation{CAS Key Laboratory of Microscale Magnetic Resonance and School of Physical Sciences, University of Science and Technology of China, Hefei 230026, China}
\affiliation{CAS Center for Excellence in Quantum Information and Quantum Physics, University of Science and Technology of China, Hefei 230026, China}
\affiliation{Hefei National Laboratory, University of Science and Technology of China, Hefei 230088, China}


\maketitle
\section{Simulation of PQGHM by a qubit}
\newcommand\ket[1]{\left | #1  \right \rangle}
The eigenstates of the Hamiltonian in the main text with the form ${\cal H}({\bf k})={\bf d}({\bf k})\cdot\boldsymbol{\sigma}$ can be treated as the eigenstates of a qubit, and can be determined by the direction of ${\bf d}({\bf k})$.
The components of GHM in ${\bf h}=(h_x,h_y,h_z)$ are explicitly given by
$h_{x}=t_{1}(1+\cos k_{1}+\cos k_{2})+t_{3}[2\cos(k_{1}-k_{2})+\cos(k_{1}+k_{2})]$,
$h_{y}=t_{1}(\sin k_{1}+\sin k_{2})+t_{3}\sin(k_{1}+k_{2})$, and
$h_{z}=2t_{2}\sin\phi[\sin k_{1}-\sin k_{2}-\sin(k_{1}-k_{2})]$.
For the Hamiltonian under periodically quenching parameters
\begin{equation}
\left(t_{3}, \phi\right)=\left\{\begin{array}{lr}
\left(+0.75,-\frac{\pi}{6}\right), & t \in\left[n T, n T+T_{1}\right) \\
\left(-0.75,-\frac{\pi}{2}\right), & t \in\left[n T+T_{1}, n T+T_{1} +T_{2}\right),
\end{array}\right.
\end{equation}
the effective Hamiltonian can be calculated by the Floquet operator.
The evolution before and after quenching can be expanded into
\begin{equation}
e^{-iT_{m}{\bf h}_{m}({\bf k})\cdot\boldsymbol{\sigma}} = \cos \left|T_{m}{\bf h}_{m} \right| \sigma_{0}-i \sin \left|T_{m} {\bf h}_{m} \right| \widehat{{\bf h}}_{m} \cdot \boldsymbol{\sigma},
\end{equation}
where $\sigma_0$ is the identity matrix, $m=1,2$ and $\widehat{{\bf h}}_{m} = {\bf h}_m/\left | {\bf h}_m \right | $ is the unit direction vector of ${\bf h}_m$.
The Floquet operator therefore can be expanded into the form using Pauli matrices:
\begin{equation}\label{eq:Uexpand}
\begin{aligned} U({\bf k}) &= e^{-iT_{2}{\bf h}_{2}({\bf k})\cdot\boldsymbol{\sigma}}e^{-iT_{1}{\bf h}_{1}({\bf k})\cdot\boldsymbol{\sigma}} \\&=\left(\cos \left|{T_{1} \bf{h}}_{1}\right| \cos \left|T_{2} {\bf{h}}_{2}\right|-\sin \left|T_{1} {\bf{h}}_{1}\right| \sin \left|T_{2} {\bf{h}}_{2}\right| \cos \alpha\right) \sigma_{0} \\ &-i \cos \left|T_{1}{\bf{h}}_{1}\right| \sin \left|T_{2} {\bf{h}}_{2}\right| \widehat{\bf{h}}_{2} \cdot \boldsymbol{\sigma}-i \cos \left|T_{2} {\bf{h}}_{2}\right| \sin \left|T_{1} {\bf{h}}_{1}\right| \widehat{\bf{h}}_{1} \cdot \boldsymbol{\sigma} \\ &+i \sin \left|T_{1} {\bf{h}}_{1}\right| \sin \left|T_{2} {\bf{h}}_{2}\right|\left(\widehat{\bf{h}}_{1} \times \widehat{\bf{h}}_{2}\right) \cdot \boldsymbol{\sigma} \end{aligned}
\end{equation}
where $\cos \alpha \equiv \widehat{\bf{h}}_{1} \cdot \widehat{\bf{h}}_{2}$.
Using quasienergy $E(\bf k)$, the Floquet operator can also be written in the form
\begin{equation}\label{eq:Uquasi}
    U({\bf k}) = e^{-i E({\bf k})\widehat{{\bf{r}}}_{U} \cdot \boldsymbol{\sigma}} = \cos[E({\bf k})]\sigma_0-i \sin[E({\bf k})]\widehat{{\bf{r}}}_{U} \cdot \boldsymbol{\sigma},
\end{equation}
where $\widehat{{\bf{r}}}_{U}$ is a unit vector.
According to eq.~\ref{eq:Uexpand} and eq.~\ref{eq:Uquasi}, $E(\bf k)$ can be calculated by
\begin{equation}
    E({\bf k}) = \pm \arccos[\left(\cos \left|{T_{1} \bf{h}}_{1}\right| \cos \left|T_{2} {\bf{h}}_{2}\right|-\sin \left|T_{1} {\bf{h}}_{1}\right| \sin \left|T_{2} {\bf{h}}_{2}\right| \cos \alpha\right)].
\end{equation}
We can calculate $\widehat{{\bf{r}}}_{U}$ which is the Bloch vector of the eigenstate using
\begin{equation}
 \widehat{{\bf{r}}}_{jU} =-\frac{\mathrm{Im} \{ \mathrm{Tr} [U({\bf k})\sigma_j] \}}{2\sin[E({\bf k})]},  
\end{equation}
where $j=x,y,z.$
The eigenstates thus can be found using the Bloch vectors.
And we can also measure the spin expectation of the qubit to obtain $\widehat{{\bf{r}}}_{U}$ and then obtain the spin texture of the Floquet band.
The pulse sequence that drives the evolution of the initial state to the eigenstate is the sequence of SWA detection introduced in subsection.~\ref{pulseSequence}.
To explore the topology of the system, we choose the eigenstate to fill one of the Floquet bands.
The Floquet operator $U({\bf k})$ also describe periodic driving of a qubit in the rotating frame.
Therefore, we can also use this evolution operator to drive the qubit to simulate the system.
To achieve this, let the evolution of each segment be consistent with that of $U({\bf k})$. The details are intodecred in the DWA detection part of subsection.~\ref{pulseSequence}.

\section{Experiment}
We detect the spin texture in the SWA experiments using three different sets of parameters: $(T_1,T_2) = (0.3,0.3)$, $(0.9,0.8)$ and $(0.9,1.2)$.
We denote these three sets of experiments as experiments A, B, and C, respectively.
This is consistent with the markers in Fig.~2(a) in the main text.
And we denote the DWA experiments in the main text as experiments D.

\subsection{Experimental setup}

Two bulk diamonds are used throughout these experiments. The one in $\left \langle 111	\right \rangle$ direction is used in experiment A. And the one in $\left \langle 100	\right \rangle$ direction is used in experiment B, C and D.

We implement the experiments by a home-built confocal microscope.
The $532$ nm laser is used for optical excitation.
The laser is focused into the diamond by an oil objective (100*O, NA 1.45).
The laser beam is released and cut off by an acousto-optic modulator (AOM).
And we use a double pass AOM system to reduce the laser leakage. 
Fluorescence counting rates range from 200 thousand to 300 thousand counts per second in these experiments.

The microwaves are generated by an arbitrary waveform generator (AWG) and then strengthened by a power amplifier(Mini Circuits ZHL-25W-63+).
We use two AWGs for different experiments (Keysight M8195A for experiment A and Keysight M8190A for experiment B, C and D).
The TTL signals are generated by a pulse generator (SpinCore PulseBlasterESR-PRO-500) to synchronize laser, microwave pulses and counter card.

\subsection{Calibration of MW amplification} \label{Calibration}
The AWG's output we used in DWA experiments has a relatively wide microwave amplitude range, therefore the power amplifier's amplification effect is not sufficiently linear.
Therefore we need to calibrate the Rabi frequency as a function of the resonant microwave amplitude.
To perform the calibration, we measure the frequency of Rabi oscillation at various microwave output amplitudes. 
Then, these data are fitted to provide a functional expression that adequately matches.
Such calibration is carried out before the DWA experiments.
Table~\ref{RabiTable} shows the data of one calibration experiment.
The microwave amplitudes in these measurements are greater than 0.05.
For data at amplitudes below 0.05 we use a linear fit to calculate the amplification away from the saturation region.
The calibration is not needed in SWA experiments because the output amplitude of the microwave is fixed and what needs to be varied is the microwave duration.
\begin{table}[]
	\begin{tabular}{l|l||l|l}
	\hline
	MW Amplitude & Rabi Frequency (MHz)   & MW Amplitude    & Rabi Frequency (MHz)   \\ \hline
    0.05         & 1.5         & 0.6          & 15.4         \\
    0.1          & 3.0         & 0.7          & 17.1         \\
    0.2          & 5.8         & 0.8          & 18.8         \\
    0.3          & 8.5         & 0.9          & 20.4         \\
    0.4          & 11.1        & 1.0          & 21.8         \\
    0.5          & 13.4        &              &              \\ \hline
	\end{tabular}

	\caption{\label{RabiTable}
	Rabi oscillation frequency for various microwave output amplitudes.
	The maximum amplitude of the microwave is defined as 1.
	The fitting function is taken as $\omega_{R} = ae^{-bA}+c$ where $\omega_{R}$ is Rabi oscillation Frequency, $A$ is the microwave amplitude, a, b, and c are the fitting parameters.
	The fitting results are $a = -36.95 (-38.52, -35.37)$, $b = 0.9092 (0.8448, 0.9736)$, and $c = 36.72 (35.04, 38.41)$.
	}
\end{table}

\subsection{Pulse Sequence}\label{pulseSequence}
In the laboratory frame, the Hamiltonian of the qubit driving by microwave is
\begin{equation}\label{PrepLab}
H^{{\rm{lab}}} = \frac{\omega _0}{2}\sigma _z + \omega_1 \cos ( {\omega _2} t +\varphi_{\rm{ini}})\sigma _x.
\end{equation}
where the first term on the right-hand side is the static component of the Hamiltonian with resonant frequency $\omega _0$,
and the second term accounts for the effect of the microwaves with Rabi frequency $\omega_1$, initial phase $\varphi_{\rm{ini}}$, and time duration $t$.
And the frequency of the microwave is set to $\omega_2$, to drive the qubit in or off resonance, depending on whether $\omega_2$ equals $\omega_0$ or not.
One can transform this Hamiltonian into the new form in the rotating frame. This transformation can be used to obtain the static Hamiltonian terms generated by the microwaves.
Under the rotating transformation $R = {e^{ - i{\omega _2} t{\sigma _z}/2}}$, the Hamiltonian in Eq.~(\ref{PrepLab}) can be transformed to the form
\begin{equation}\label{PrepRotDWA}
	\begin{aligned}
	H^{{\rm{rot}}} &= {R^\dag }H^{{\rm{lab}}}R - i{R^\dag }\frac{d}{{dt}}R \\
	&= \frac{\omega _1}{2} ( \sigma _x\cos \varphi_{\rm{ini}} +\sigma _y\sin \varphi_{\rm{ini}}) + \frac{\omega _0 - \omega _2}{2} \sigma _z,
	\end{aligned}
\end{equation}
where the second equality relies on the rotating wave approximation which drops the faster rotating term in the rotating frame.

In SWA experiments, qubit is initialized to $\left| {\psi _{{\rm{ini}}}^{{\rm{rot}}}} \right\rangle = \left| 0 \right\rangle$ and then drived by the microwave pulse with frequency $\omega_2$ which is set equal to $\omega_0$.
So the Hamiltonian of the qubit is
\begin{equation}\label{PrepLabSWA}
	H_{\rm{SWA}}^{{\rm{lab}}} = \frac{\omega _0}{2}\sigma _z + \omega_1 \cos ( {\omega _0} t +\varphi_{\rm{ini}})\sigma _x.
\end{equation}
According to Eq.~(\ref{PrepRotDWA}), under the rotating transformation $R = {e^{ - i{\omega _0} t{\sigma _z}/2}}$,
the Hamiltonian in Eq.~(\ref{PrepLabSWA}) can be transformed to the form
\begin{equation}\label{PrepRotSWA}
	\begin{aligned}
	H_{\rm{SWA}}^{{\rm{rot}}}
	&= \frac{\omega _1}{2} ( \sigma _x\cos \varphi_{\rm{ini}} +\sigma _y\sin \varphi_{\rm{ini}}),
	\end{aligned}
\end{equation}

For each k, we calculate the vector ${\bf d}({\bf k})$ from the Floquet effective Hamiltonian ${\cal H}({\bf k})={\bf d}({\bf k})\cdot\boldsymbol{\sigma}$.
And then we set $\varphi_{\rm{ini}}=\arg(d_x+id_y) + \pi/2$ and the pulse duration $\tau_{\rm{pul}} = \arccos (d_z/\left | {\bf d} \right | ) / \omega_1$.
Experimentally, Rabi frequency $\omega_1$ is obtained through the Rabi oscillation experiment and the calibration process in subsection.~\ref{Calibration}.
So the qubit evolves to
\begin{equation}\label{finalStateRot}
	\left| {\psi _{{\rm{fin}}}^{{\rm{\rm{SWA}}}}} \right\rangle = e^{-iH_{{\rm{SWA}}}^{{\rm{rot}}}\tau_{\rm{pul}}} \left| {\psi _{{\rm{ini}}}^{{\rm{SWA}}}} \right\rangle = |u_{s}({\bf k})\rangle,
\end{equation}
where $|u_{s}({\bf k})$ is one of the Floquet eigenstates of ${\cal H}({\bf k})$ in the main text.

Eq.~(\ref{PrepRotSWA}) provides the basis for us to construct a Hamiltonian involved $\sigma _x$ and $\sigma _y$.
In DWA experiments, $\sigma _z$ is also required, and this can be realized by driving off-resonance to get such term in Eq.~(\ref{PrepRotDWA}).
For convenience, the qubit is initialized to $\left| {\psi _{{\rm{ini}}}^{{\rm{DWA}}}} \right\rangle = \left| 0 \right\rangle$.
The qubit undergoes 64 periods of evolution, each of which consists of two parts, marked by $T_1$ and $T_2$ in Fig.~4 and and illustrated in Eq.~(1) in the main text.

In each part, the Hamiltonian can be expressed in the rotating frame as Eq.~(\ref{PrepRotDWA}),
and we set $\varphi_{\rm{ini}}=\arg(h_x+ih_y)$ and let $\omega_1$, $\omega_2$ and pulse duration $\tau_m$satisfy
\begin{equation}\label{freDemands}
	\frac{1}{2}(\omega_1 \cos \varphi_{\rm{ini}},\ \omega_1\sin \varphi_{\rm{ini}}, \ \omega_0 - \omega_2) \cdot \tau_m = {\bf h}_m T_m,
\end{equation}
where $m=1,2$.

\newcommand\DWAevo[1]{ e^{-iH_{{T_{#1}}}^{{\rm{rot}}}\tau_{#1}} }

Then the qubit evolves to
\begin{equation}\label{finalStateRot}
	\left| {\psi _{{\rm{fin}}}^{{\rm{\rm{DWA}}}}} \right\rangle = (\DWAevo{2}\DWAevo{1})^{N} \left| {\psi _{{\rm{ini}}}^{{\rm{DWA}}}} \right\rangle,
\end{equation}
where $N$ is the number of driving periods.
In such situation, the qubit is sensitive to static magnetic noise. Therefore, in some sequences, we amplify the microwave amplitude and shorten the duration to maintain the total accumulated phase while suppressing noise effects.

As shown in Fig.~3(e) in the main text, the spin state is read out during the latter laser pulse and there are two counting windows.
The total photon count recorded by the first (second) window during these iterations is regarded as signal (reference) and denoted by $s$ ($r$).
The raw experimental data is $x=(s-r)/r$. 
To normalize the data, a conventional Rabi oscillation is performed. 
We fit the raw data of the Rabi oscillation using the function $x=x_0+a\cos(\omega_{\rm R} t+\varphi)$, and then use the range of the function to normalize the experimental data as $x_{\rm{n}}=(x-x_0)/(2a)+1/2$.
The data thus normalized represent the probability of measuring $\left| {{m_s} =  0} \right\rangle$.
And then the data are transformed to the $\left \langle \sigma _z \right \rangle$ of the qubit using the linear function $y_{\rm{n}} = 2x_{\rm{n}}-1$.

\subsection{SWAs distribution}

\begin{figure}
	\includegraphics[width=1\columnwidth]{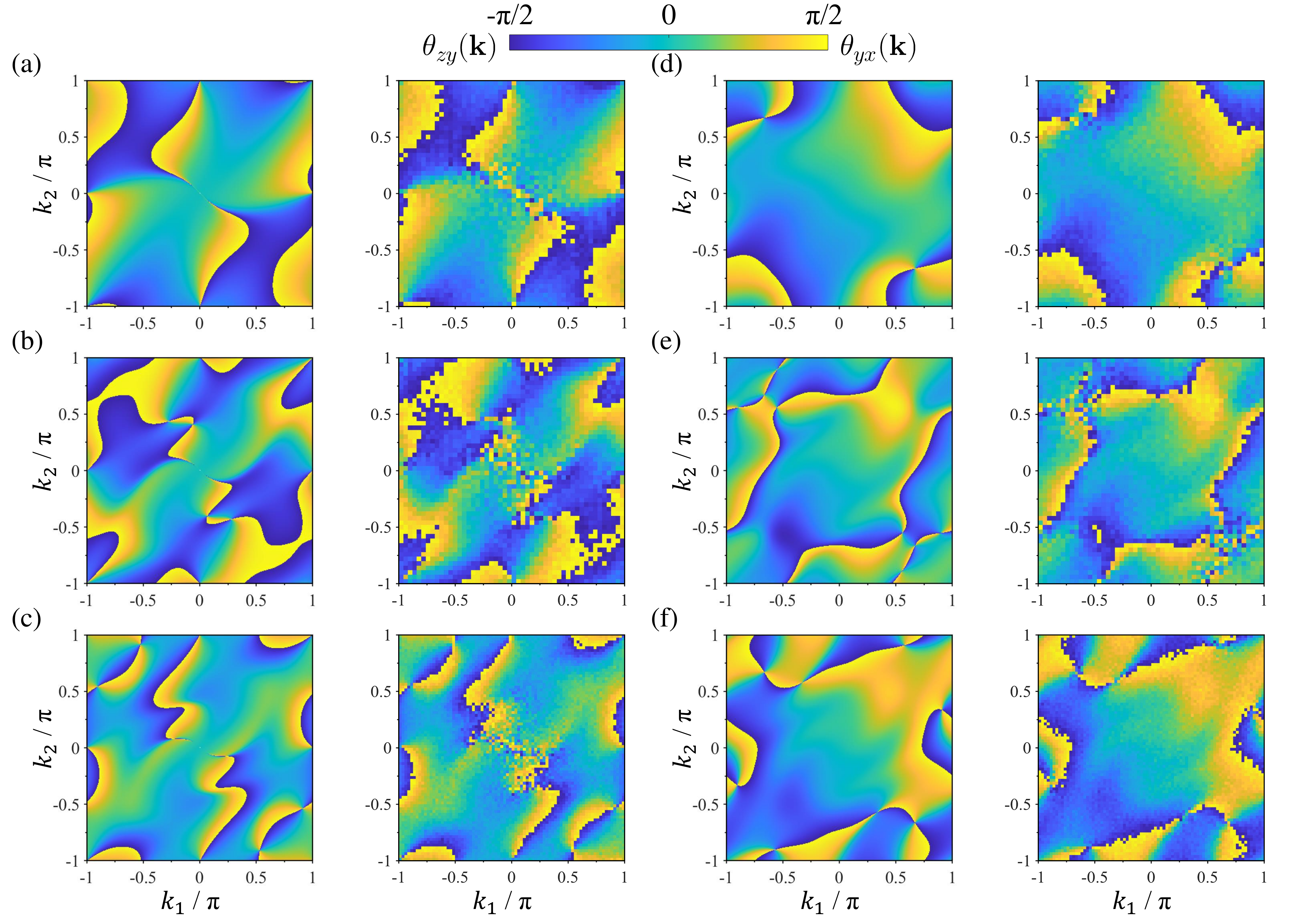}
	\caption{
Static winding angle ($\theta_{zy}$ and $\theta_{yx}$).
(a)--(f) show the distribution of SWAs in ${\bf k}$-space. All the left and right panels are obtained from theory and experiment respectively.
The parameters in (a), (b) and (c) are respectively set as $(T_1,T_2) = (0.3,0.3)$, $(0.9,0.8)$ and $(0.9,1.2)$. Other parameters are the same as those in Fig.~2 in the main text.
The parameters in (d), (e) and (f) are the same as those in (a), (b), and (c), respectively.
}
\label{SWA_SM}
\end{figure}

In Fig.~3 in the main text we show the $\theta_{xz}$ distribution in ${\bf k}$-space, and here we show the $\theta_{zy}$ and $\theta_{yx}$ distributions in Fig.~\ref{SWA_SM}.
Data are acquired at the same time as the experiments in the main text.
Some singularity points in the distribution are very close to each other, so to calculate SWA, the sampling around these points needs to be denser.

\subsection{DWAs distribution}
\begin{figure}
	\includegraphics[width=1\columnwidth]{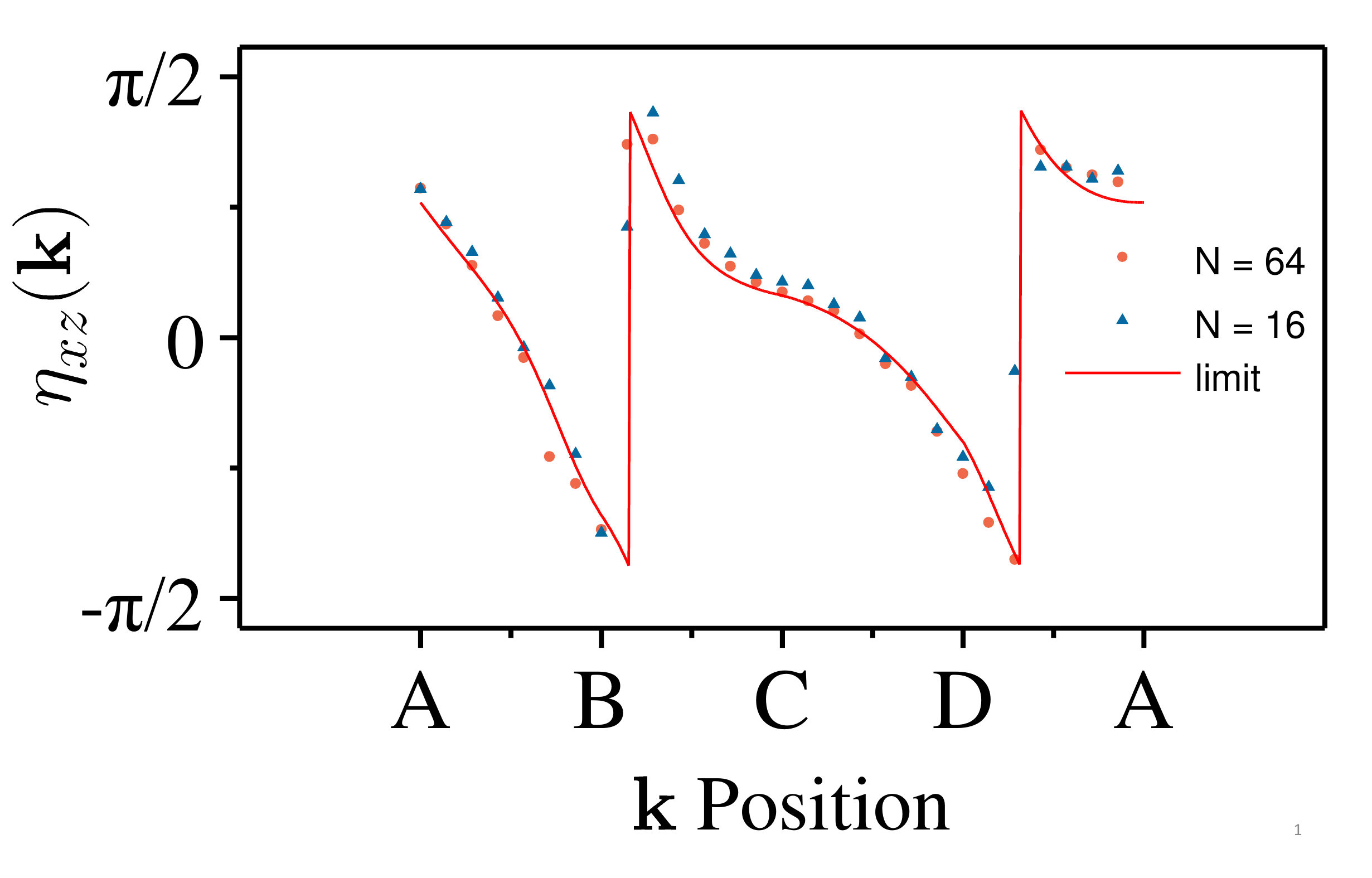}
	\caption{
		DWA distribution on a closed loop. The circles and triangles represent the data of $N=64$ and $N=16$ driving periods, respectively.
		The red curve represents theoretical results of the $N\rightarrow\infty$ limit.
		The loop and the data of $N=64$ are the same as those shown in Fig.~4(d) in the main text.
		}
\label{DWA_SM}
\end{figure}

Fig.~\ref{DWA_SM} shows the experimental values of DWAs under different driving periods $N$.
Here in the limit $N\rightarrow\infty$, $\eta_{jl}({\bf k})=\theta_{jl}({\bf k})$.
One can see that in Fig.~\ref{DWA_SM}, the two sets of experimental data are fairly close, except near the discontinuity point, where the DWA has a jump between $-\pi/2$ and $\pi/2$.
Increasing $N$ greatly reduces the difference between DWA and SWA near the discontinuity point, and suppresses the error of calculating the winding number.

\nocite{*}